\documentclass[aps,prl,reprint,superscriptaddress,footinbib,longbibliography]{revtex4-1}

\usepackage{hyperref} 
\usepackage[centertags,sumlimits,intlimits,namelimits,reqno]{amsmath}
\usepackage{graphics,graphicx,epsfig,bbm}
\usepackage{amsfonts,amssymb,amscd,latexsym,color,gensymb}
\usepackage{times}
\usepackage{multirow}

\newcommand{\ketbra}[2]{|{#1}\rangle\!\langle{#2}|}

\newcommand{\op}[1]{{\hat{#1}}}
\newcommand{\sop}[1]{{\mathcal{#1}}}

\newcommand{\ot}{\otimes}

\DeclareMathAlphabet{\matheu}{U}{eus}{m}{n}

\newcommand{\ket}[1]{|{#1}\rangle}

\newcommand{\expect}[2]{\langle {#1} \rangle_{#2}}

\DeclareMathOperator{\tr}{tr}

\newcommand{\U}{\sop{U}}

\newcommand{\tU}{\tilde{\sop{U}}}

\newcommand{\MI}{\op{I}\otimes \op{Y}\otimes \op{I}\otimes \op{X}\otimes \op{Z}}
\newcommand{\MO}{\op{Z}\otimes \op{Z}\otimes \op{I}\otimes \op{Y}\otimes \op{X}}

\begin{document}

\title{Practical experimental certification of computational 
quantum gates via twirling}

\author{Osama Moussa}
\email{omoussa@iqc.ca}
\affiliation{
Institute for Quantum Computing
and 
Department of Physics and Astronomy, 
University of Waterloo, Waterloo, ON, N2L 3G1, Canada.
}

\author{Marcus P. da Silva}
\email{msilva@bbn.com}
\affiliation{
Depart{\'e}ment de Physique, 
Universit{\'e} de Sherbrooke, Sherbrooke, QC, J1K 2R1, Canada.
}
\affiliation{
Raytheon BBN Technologies,
Disruptive Information Processing Technologies Group,
Cambridge, MA, 02138, USA
}

\author{Colm A. Ryan}
\affiliation{
Institute for Quantum Computing
and 
Department of Physics and Astronomy, 
University of Waterloo, Waterloo, ON, N2L 3G1, Canada.
}
\affiliation{
Raytheon BBN Technologies,
Disruptive Information Processing Technologies Group,
Cambridge, MA, 02138, USA
}

\author{Raymond Laflamme}
\email{laflamme@iqc.ca}
\affiliation{
Institute for Quantum Computing
and 
Department of Physics and Astronomy, 
University of Waterloo, Waterloo, ON, N2L 3G1, Canada.
}
\affiliation{
Perimeter Institute for Theoretical Physics, Waterloo, ON, N2J 2W9, Canada.
}

\begin{abstract} 
  Due to the technical difficulty of building large quantum computers,
  it is important to be able to estimate how faithful a given
  implementation is to an ideal quantum computer. The common approach
  of completely characterizing the computation process via quantum
  process tomography requires an exponential amount of resources, and
  thus is not practical even for relatively small devices. We solve
  this problem by demonstrating that twirling experiments previously
  used to characterize the average fidelity of quantum memories
  efficiently can be easily adapted to estimate the average fidelity
  of the experimental implementation of important quantum computation processes, 
  such as unitaries in the Clifford group, in a practical
  and efficient manner with applicability in current quantum
  devices. Using this procedure, we demonstrate state-of-the-art
  coherent control of an ensemble of magnetic moments of nuclear spins
  in a single crystal solid by implementing the encoding operation
  for a 3 qubit code with only a 1\% degradation in average fidelity 
  discounting preparation and measurement errors. We also highlight one of the advances
that was instrumental in achieving such high fidelity control.
\end{abstract}

\maketitle

\textbf{Introduction --} Due to the technical challenges of building
quantum computers, only small building blocks of such devices have
been demonstrated so far in a number of different physical systems. In
order to quantify how closely these demonstrations come to the desired
ideal operations, the experiments are fully characterized via {\em
  quantum process tomography} (QPT)~\cite{PCZ97,CM97}, and, often, the {\em
  average fidelity}~\cite{BOS+02,Nie02} between the experiment and the
ideal operator is calculated from the description of the estimated
process. The main drawback of this approach is that quantum process
tomography fundamentally requires an exponential number of
experiments. Moreover, the classical post-processing of the data is
non-trivial, as the raw experimental data does not lead to a physical
description, and approaches such as maximum likelihood or Bayesian
estimation on an exponentially large parameter space are needed to
find the most appropriate physical description~\cite{Blu10}. Therefore
approaches based on QPT to estimate the average fidelity are not practical, and
cannot be reasonably expected to be used even in systems that are only
moderately larger than the current experimental state of the art.
Here we solve this problem by showing that, for an important
class of quantum operations, the average fidelity can be estimated
efficiently, requiring a number of experiments which is independent of
the system size. This new proposal is also practical, and enables the
demonstration of processes which would not have been
possible due to the complexity of QPT.
We use this protocol to demonstrate state-of-the-art
coherent control of the magnetic moments of an ensemble of nuclear
spins in a single crystal solid, and highlight one of the advances
that was instrumental in achieving such high fidelity control.

\begin{figure}
\includegraphics[width=.5\textwidth]{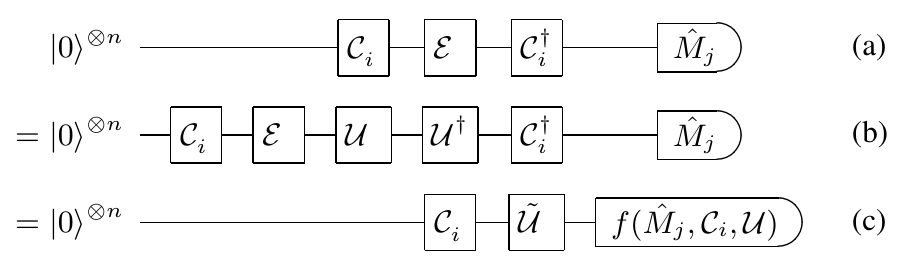}
\caption{The relationship between (a) comparing a physical process
  ${\mathcal E}$ to the identity, and (c) comparing a physical process
  $\tilde{\mathcal U}={\mathcal U}\circ{\mathcal E}$ to a unitary
  ${\mathcal U}$ can be seen by (b) the appropriate insertion of the
  identity ${\mathcal U}^\dagger\circ{\mathcal U}$. When $\sop{U}$ and
  all $\sop{C}_i$ are elements of the Clifford group, and $\op{M}_j$
  is a Pauli operator, $f(\op{M}_j,\sop{C}_i,\sop{U})$ is also a Pauli
  operator.\label{fig:circuit}}
\end{figure}

{\bf Twirling --} It has been recently shown that if one wishes to
compare an experimental implementation of a quantum process to the
identity process ({\em i.e.} a process where the system state remains
unchanged, such as in the case of ideal quantum memories), then it is
possible to estimate the average fidelity via a technique known as
{\em twirling}~\cite{EAZ05,DCEL09,ESM+07,Silva08,LBP+10}, with a number of
experiments which depends only on the desired accuracy of the
estimate, not on the system size --- moreover, these experiments are
simple to implement, requiring only local operations and
measurements~\cite{ESM+07}.  The twirling procedure consists of
applying a random unitary before the process to be characterized,
followed by the inverse of this randomly chosen unitary.  When these
unitaries obey certain symmetry properties, the resulting invariant
information about the noise under this symmetry can be extracted by
repeating the experiment with different random choices.  For example,
if the twirling gates are random permutations followed by tensor
products of single-qubit Cliffords then information about the weights
of the noise terms can be determined~\cite{ESM+07}.  The schematic for
such an experiment is depicted in Fig.~\ref{fig:circuit}(a), where the
process we would like to compare to the identity process, $\mathcal
E$, is conjugated with $\mathcal C_i$, an appropriately chosen
randomizing local operation, and $\op{M}_j$ is the measurement of the
parity of a subset of qubits in the computational basis.

For more general processes, in order to compare a given process to a
desired unitary evolution one could in principle apply the physical
process under consideration, and then apply the inverse of the unitary
evolution we would like to compare it against, finally measuring the
overlap between the initial state and the resulting state for a set of
initial states --- this is, in essence, the definition of the average
fidelity.  The obstacle to implementing such a protocol is that often
one is attempting to demonstrate or {\em certify} the implementation
of a unitary, and a noiseless implementation of its inverse cannot be
assumed to be available. One way~\cite{EAZ05,KLR+08,MGE11} to address
this problem is to estimate the average fidelity over a set of quantum
processes that form a group by considering random sequences of such
processes chosen to result in the identity process --- examples of
such sets include the group of all unitary processes, as well as the
Clifford group~\cite{GC99}.  Such motion-reversal benchmarking schemes
suffer from two shortcommings: they apply only to noise satisfying
certain strength conditions~\cite{MGE11}, and they only provide
information about the average over a set of processes instead of
specific information about a particular process. While this
information is useful, what is experimentally most useful is to
diagnose coherent control implementation errors, which, in our
experience, are highly process-dependent.  Therefore, it is critical
for the experimentalist to be able to characterize a particular
process.

\textbf{Certification procedure --} The result we report here, which
side-steps many of the shortcomings listed above, is that the average
fidelity between any physical process on multiple qubits and any
particular element of the Clifford group can be estimated efficiently
by a simple modification to the twirling protocol, leading to the same
favourable scaling as experiments which compare a physical process to
the identity. If we define ${\mathcal U}$ to be the desired element of
the Clifford group, then the noisy implementation
$\tilde{\mathcal{U}}$ can be thought of as some noisy process
${\mathcal E}$ followed by the application of ${\mathcal U}$,
i.e. $\tilde{\mathcal U} = {\mathcal U} \circ {\mathcal E}$.
Unitaries in the Clifford group include operations needed to encode
and decode quantum information to protect it from noise~\cite{Got99}
--- in current approaches to fault-tolerance these operations comprise
the vast majority of (if not all) operations.  Clifford group
operations can also be used to achieve universal fault-tolerant
quantum computation with the aid of especially prepared resource
states~\cite{GC99,BK05}, so these operations are of great importance
and utility for quantum computation.

In order to see why the average fidelity can be estimated efficiently
for these operations, consider Fig.~\ref{fig:circuit}(b), which
modifies the original twirling protocol~\cite{ESM+07} by inserting the
identity process --- in this case written as ${\mathcal
  U}^\dagger\circ{\mathcal U}$. One can in principle combine all
processes after the first application of ${\mathcal U}$ in (b) into a
new measurement. For a general unitary process this new measurement
will be as hard to implement as performing the process ${\mathcal U}$
itself. However, if ${\mathcal U}$ is an element of the Clifford
group~\footnote{Other operations can also be certified efficiently in
  this manner. The fundamental requirement is that the new measurement
  $f(\op{M}_j,\sop{C}_i,\sop{U})$ be a measurement that is easily
  implemented with high fidelity.}, this results in the measurement of
the parity of a different set of qubits in a different local basis (or
equivalently, the measurement of a different Pauli
operator~\cite{GC99}) which can be precomputed efficiently given
${\mathcal U}$, the local randomizing Clifford operation ${\mathcal
  C}_i$, and the original measurement $\op{M}_j$, as depicted in
Fig.~\ref{fig:circuit}(c), where
$f(\op{M}_j,\sop{C}_i,\sop{U})=\U(C_i\op{M}_{j}C_i^\dagger)$. In
essence, the protocol in Fig.~\ref{fig:circuit}(c) is the experiment,
but the data is analyzed according to Fig.~\ref{fig:circuit}(a) as
described in~\cite{ESM+07,Silva08}, which separates the noise
${\mathcal E}$ from the unitary ${\mathcal U}$.

As the parity measurement is equivalent to local measurement followed
by simple data post-processing, and the initial states required are
product states locally equivalent to the all-zeros state, it is
important that precise local operation be available.  In other words,
the problem of implementing the inverse of a multibody Clifford
unitary ${\mathcal U}$ can be translated into the problem of
implementing classical data processing and local (single-body) quantum
operations reliably. These operations are often readily available at
high fidelities, as randomizing benchmarking results have
demonstrated~\cite{KLR+08,RLL09,CGT+09}.  Thus the average fidelity of
any implementation of a Clifford group operation can be estimated
using a number of experiments that depends only on the desired
accuracy, as is the case for twirling experiments with quantum
memories~\cite{ESM+07}.

Due to this connection to twirling protocols, our proposal also
enables the estimation of other parameters beyond the gate fidelity,
such as the probability of errors of a given weight. Recent proposals
for Monte Carlo estimation of state and gate fidelity have the same
scaling as the protocol we describe here (in the case of Clifford
gates)~\cite{FL11,SLP11}. However, the probability of errors of a
given weight are not natural parameters to be considered in the Monte
Carlo sampling proposals, demonstrating the advantage of considering
twirling protocols in this context.  The simplicity of the experiments
also shows that our proposal is of practical significance in the
benchmarking of these important operations. Moreover, because the
estimation of the average fidelity in the twirling protocol
corresponds to the estimation of the probability of no errors having
occurred (a single parameter that is accessible with an accuracy that
does not depend on the the number of qubits~\cite{ESM+07}), Bayesian
estimation of such a probability is straightforward, as is the
calculation of uncertainties associated with these estimates.

\textbf{Experiment -- } A common task for an experimentalist is to optimize and tweak the performance of a particular gate on the system.  The experimenter has many potential knobs to adjust and he/she needs a reliable robust method for certifying whether any changes actually improved the performance.  A trivial example is calibrating the power of a pulse but here we demonstrate how we can easily quantify the improvement from more subtle and sophisticated control improvements.  

Building on the success of liquid-state NMR as
a test bed of QIP ideas, Solid-state NMR systems
offer~\cite{CLK+00,BMR+06} intrinsically larger couplings, longer
coherence times, the ability to pump entropy out of the system of
interest into a spin bath~\cite{BMR+05,RMB+08} and the potential for much higher initial
polarizations. This comes at the cost of a more complicated internal
Hamiltonian, which makes the system harder to control in practice.

\begin{figure}
\includegraphics[width=.5\textwidth]{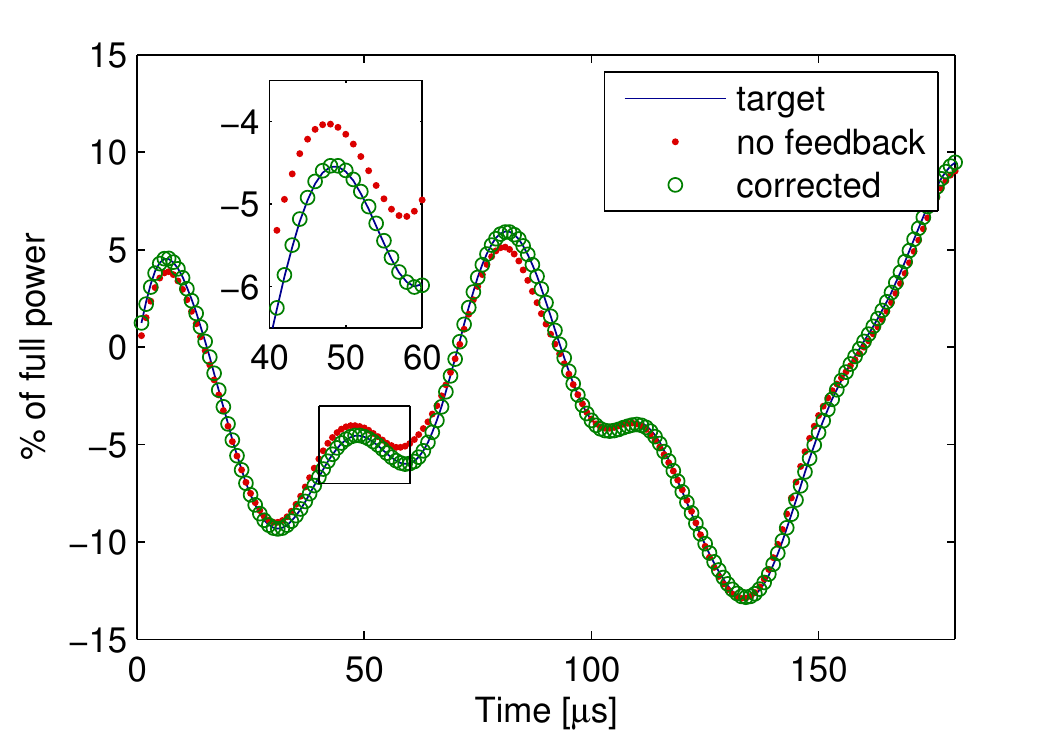}
\caption{\label{figfixing} Portion of a typical pulse shape showing
  (in solid blue) the designed target shape, (in red dots) the initial
  attempt at implementing the pulse including systematic imperfections
  due to nonlinearities in pulse generation and amplification as well
  as finite bandwidth effects from the probe's resonant circuit, and
  (in green circles) the corrected shape after the feedback
  protocol. Full power corresponds to nutation frequency of 80kHz.}
\end{figure}

Methods inspired by optimal control theory have been successful in
aiding pulse design for small systems. However, for these pulses to
achieve the designed fidelity, it is important that the implemented
control fields at the sample match the designed ones. That is to say,
any systematic deviations from the designed pulses, due to the finite
bandwidth of the resonant probe circuit or the non-linearities in the
pulse generators and amplifiers, need to be accounted for or
rectified.  To this end, a feedback system can be employed to correct
for these systematic imperfections~\cite{WHE+04}. We use an antenna to
measure the fields in the vicinity of the sample, then this data is
fed back for comparison with the target pulse, and a new pulse form
that attempts to compensate for the imperfections is computed and sent
back to the signal generation unit. This loop is repeated a number of
times to reach a satisfactory pulse
form~\cite{Ryan09,Moussa10}. Fig.~\ref{figfixing}~ shows a typical
example of the measured pulse forms of the initial and corrected
attempts to match a target pulse shape. The development of this
feedback pulse rectification protocol has led to a great improvement
in the fidelity of coherent control of nuclear spins in the solid
state -- the certification scheme is used herein to demonstrate and
quantify the typical improvement in fidelity resulting from using the
feedback system.

The specific computational register under investigation is an ensemble
of molecular nuclear spins in a macroscopic single crystal of Malonic
Acid (C$_3$H$_4$O$_4$). A small fraction ($\sim 3\%$) of the molecules
are triply labeled with (spin-$\tfrac{1}{2}$) $^{13}$C to form an
ensemble of 3-qubit processor molecules, spatially buffered from one
another by molecules of the same compound but with natural abundance
($\sim1\%$) carbon nuclei. During computation, the processors are
decoupled from the 100\% abundant spin-$\tfrac{1}{2}$ protons in the
crystal by applying a decoupling pulse sequence to the protons.

\begin{figure}
\hspace{-.4in}
\begin{minipage}{0.4\linewidth}
\includegraphics[scale=.18]{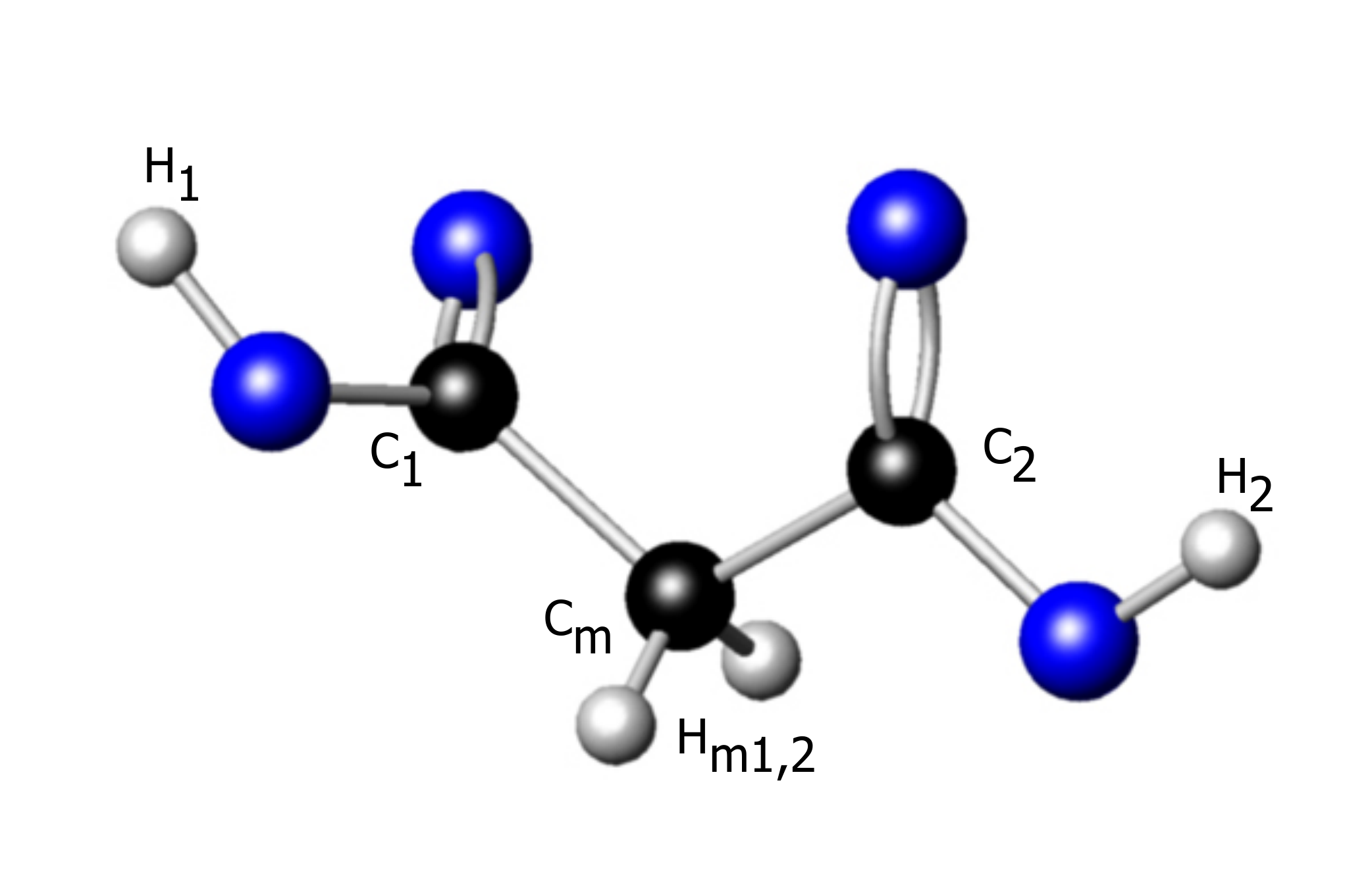}
\end{minipage}
\hspace{0.5cm}
\begin{minipage}{0.35\linewidth}
 \begin{small}
\begin{tabular}{|c|c|c|c|}
\hline
kHz & C$_1$ & C$_2$ & C$_m$\\
\hline
C$_1$ & 6.380 & 0.297 & 0.780\\
\hline
C$_2$ & -0.025 & -1.533 & 1.050\\
\hline
C$_m$ & 0.071 & 0.042 & -5.650\\
\hline
\end{tabular}
\end{small}
\end{minipage}\\
\includegraphics[width=2.8in]{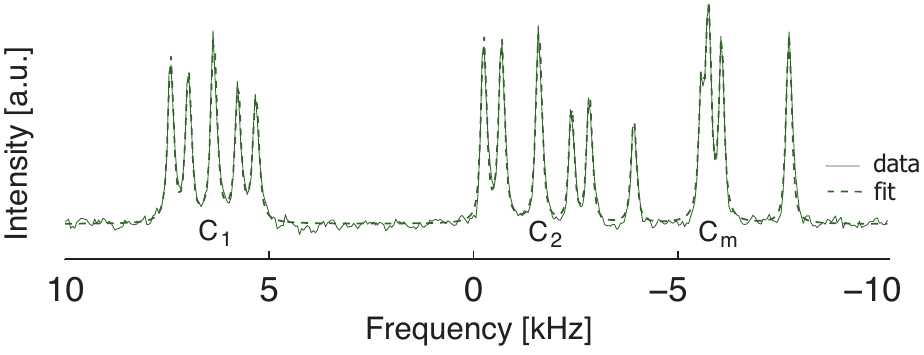}
\caption{\label{figmalonic} Malonic acid (C$_3$H$_4$O$_4$) molecule
  and Hamiltonian parameters (all values in kHz).  Elements along the
  diagonal represent chemical shifts, $\omega_i$, with respect to the
  transmitter frequency (with the Hamiltonian $\sum_i \pi \omega_i
  \op{Z}_i$). Above the diagonal are dipolar coupling constants,
  $\sum_{i<j} \pi D_{i,j} (2\ \op{Z}_i\op{Z}_j - \op{X}_i\op{X}_j -
  \op{Y}_i\op{Y}_j$), and below the diagonal are J coupling constants,
  $\sum_{i<j} \tfrac{\pi}{2} J_{i,j} (\op{Z}_i\op{Z}_j +
  \op{X}_i\op{X}_j + \op{Y}_i\op{Y}_j$).  An accurate natural
  Hamiltonian is necessary for high fidelity control and is obtained
  from precise spectral fitting of (also shown) a proton-decoupled
  $^{13}$C spectrum following polarization-transfer from the abundant
  protons. The central peak in each quintuplet is due to natural
  abundance $^{13}$C nuclei present in the crystal at $\sim 1\%$. (for
  more details see~\cite{BMR+06,RMB+08} and references therein.) }
\end{figure}

The experiments were performed at room temperature in a static field
of $7.1$T using a purpose-built NMR probe. Shown in
Fig.~\ref{figmalonic} is a proton-decoupled $^{13}$C spectrum,
following polarization-transfer from the abundant protons, for the
particular orientation of the crystal used in this experiment. A
precise spectral fit gives the Hamiltonian parameters (listed in the
inset table in Fig.~\ref{figmalonic}), as well as the free-induction
dephasing times, $T_2^*$, for the various transitions; these average
at $\sim2$ms. The dominant contribution to $T_2^*$ is Zeeman-shift
dispersion, which is largely refocused by the control pulses, leading
to effective dephasing times much larger than $T_2^*$~\cite{BMR+06}.
The carbon control pulses are numerically optimized to implement the
required unitary gates using the Gradient Ascent Pulse Engineering
(GRAPE)~\cite{KRK+05} algorithm, and are typically
designed~\cite{RNL+08} to have an average Hilbert-Schmidt fidelity of
$99.8\%$ over appropriate distributions of Zeeman-shift dispersion and
control-fields inhomogeneity.

\begin{figure}[h]
\includegraphics[scale=.85]{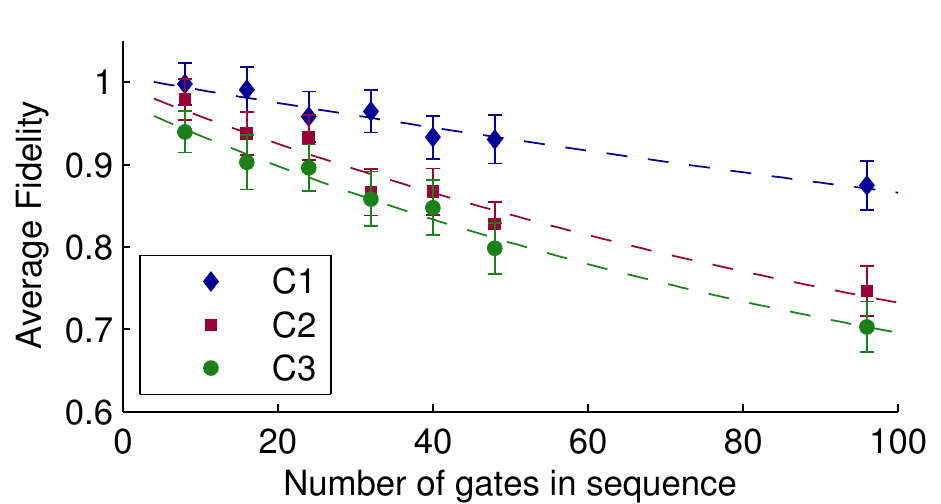}
\caption{\label{figqbm} Randomized benchmarking of single qubit
  $\tfrac{\pi}{2}$ rotations required for state preparation and
  measurement -- shown is the average fidelity decay of randomized
  sequences of $\tfrac{\pi}{2}$ pulses on each of the three
  qubits. Each data point is the average fidelity of 24
  sequences. Fitting the data to~\cite{MGE11} $\log(F-\tfrac{1}{2}) =
  \log A_0 + m \log p\,,$ we extract an average error per gate of
  $1.6\pm0.4\times 10^{-3}$ for C$_1$ (blue diamonds),
  $3.8\pm0.7\times 10^{-3}$ for C$_2$ (red squares), and
  $4.4\pm0.6\times 10^{-3}$ for C$_3$ (C$_m$) (green circles).}
\end{figure}
\vspace{-.1in}

\begin{figure*}[t]
\includegraphics[scale=1]{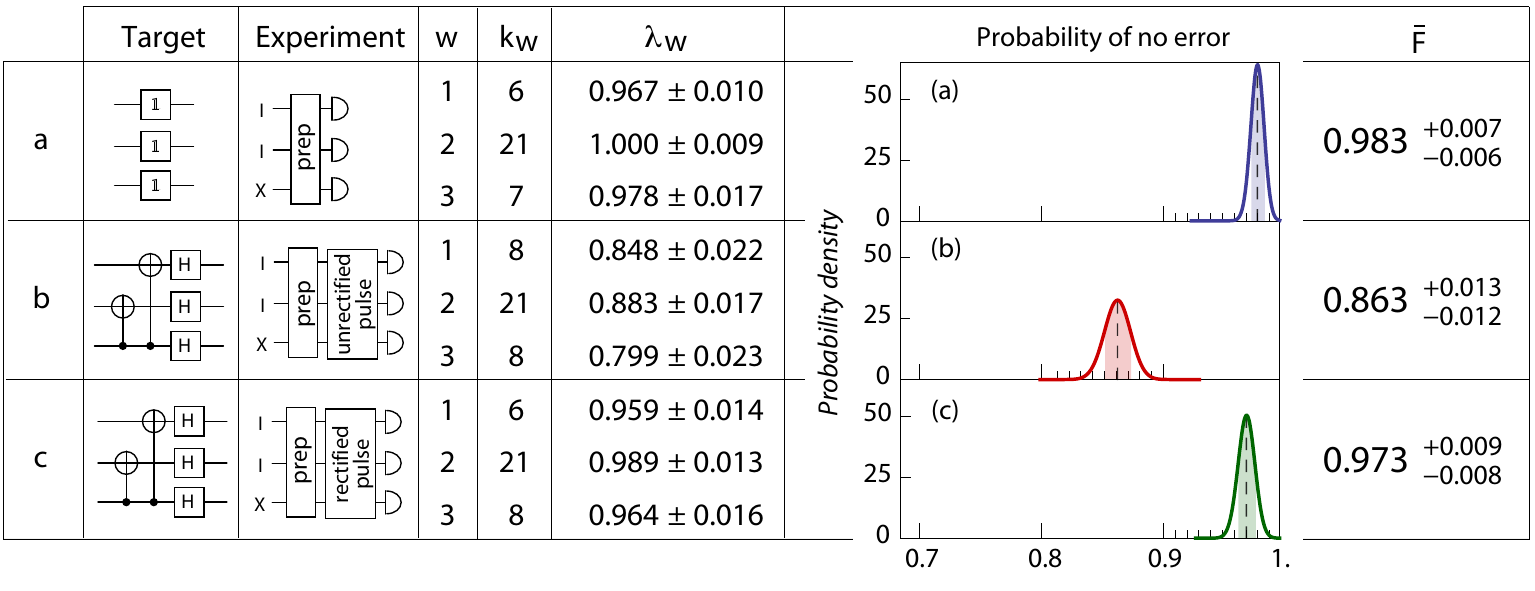}
\caption{\label{figfidelity} Summary of the experimental parameters
  and results for the three sets of certification experiments -- the
  \emph{Target} column shows the quantum circuit representation of the
  ideal process; the \emph{Experiment} column represents the
  experimental setup to certify the corresponding implementation,
  including state preparation and measurement using local readout
  pulses as described in the text; $k_w$ and $\lambda_w$ are,
  respectively, the number of performed experiments, and the average
  surviving polarization, partitioned by the Pauli-weight, $w$, of the
  input preparation (for more details, see Appendix A.) Shown also are the Bayesian estimated
  probability density functions for the probability of no error in the
  experimental implementation of the target gate, as well as the
  estimated average fidelity. The three sets of experiments are (a)
  State preparation and measurement compared to the Identity operation
  - this can be thought of as a calibration for the certification
  procedure; (b) the target is the encoding operation for the 3-qubit
  phase quantum error correcting code, and the experimental
  implementation is a numerically designed pulse using GRAPE; and (c)
  is the same as (b) but the pulse is corrected for implementation
  errors using the feedback procedure described in the text. }
\end{figure*}
\vspace{.1in}

\textbf{Preparation and measurement --} The first step in the initial
preparation procedure for all experiments described below is a
selective polarization transfer from one of the methylene protons
(H$_{m_1}$) to the methylene carbon (C$_m$). This is realized using a
short~\cite{MKB+74} Hartman-Hahn cross-polarization
sequence~\cite{HH62} after tipping the proton polarization to the
transverse plane, and is sufficient because the coupling strength
between these two nuclei is more than an order of magnitude larger
than any other coupling. The state of the three carbon nuclei after
this polarization can be described as $\rho_i = \op{I}^{\ot 3} +
\alpha \op{IIX}$, where $\alpha$ quantifies the amount of polarization
transferred from the proton, and is on the order of $\sim 10^{-5}$ for
protons in 7.1T at room temperature. A free induction decay is
collected for this initial state to establish a reference for
$\alpha$, against which all subsequent experiments are
compared. Simple coherence-transfer pulses can then be used to prepare
all states of the form $\rho_w = \op{I}^{\ot 3} + \alpha \op{X}^{\ot
  w} \op{I}^{\ot 3-w} $, and their permutations over 3 qubits, for $w=
1, 2, 3$. From these states, pulses realizing single qubit
$\frac{\pi}{2}$ rotations are all that is required for preparing a
state with non-zero projection on any arbitrary 3-qubit Pauli
operator. The same set of pulses are sufficient to transform any
output state into an observable in an NMR experiment.

These single-qubit $\frac{\pi}{2}$ rotations can be realized with very
high fidelity, which we now demonstrate using single qubit randomized
benchmarking~\cite{MGE11} on each of the qubits -- the average
fidelity of randomized sequences of pulses that compose to the
Identity is measured for varying sequence lengths, and assuming that
the implementation errors do not depend on which gate is being
applied, the average fidelity decay is fit to~\cite{MGE11}:
$F= A_0 p^m + B_0\,,$
where $m$ is the sequence length, $A_0$ and $B_0$ encompass
initialization and measurement errors, and $p$ is a parameter related
to the average error per gate, $r=\tfrac{1-p}{2}$. Assuming the gate
errors are unital, we set $B_0=\tfrac{1}{2}$.  In Fig.~\ref{figqbm},
the average fidelity of 24 sequences each for up to 96 pulses per
sequence is plotted, and the average error per single qubit
$\frac{\pi}{2}$ pulse is estimated to be less than $0.5\%$.  These
$400\mu s$ pulses are designed to selectively rotate the target qubit
while not affecting the others. As seen in Fig.~\ref{figmalonic},
the C$_1$ transitions are well separated in frequency from the other
transitions, which explains the ability to selectively rotate that
spin with better fidelity for pulses with the same length.

Furthermore, to get an estimate of the average combined fidelity of
the state preparation and measurement processes, we certify the
\emph{do nothing} operation against the target Identity evolution. The
results are summarized in Fig.~\ref{figfidelity}a.

\textbf{Certifying the 3-qubit encoding --} Next, we choose to certify
the ($1.5\ ms$) pulse~\cite{MBR+11} designed to perform the encoding
operation of the phase variant of the 3 qubit quantum error correcting
code against the ideal gate~\cite{Bra+96}, which is a 3-qubit Clifford
gate that decomposes to two CNOTs followed by transversal single-qubit
Hadamards. As shown in Fig.~\ref{figfidelity}, the average fidelity
of the implemented pulse, before and after rectification---including
preparation and measurement errors---is estimated to be 86.3\% and
97.3\%, respectively.  Under an assumption that the errors from
preparation and measurement are factorable, we estimate the average
fidelity of the rectified implementation to be 99\%.

\textbf{Discussion --} We have shown how it is possible to certify
individual Clifford group operations efficiently using a modified
twirling protocol. As an illustrative example, we demonstrated the
certification of the encoding operation for a 3 qubit error correction
code, and the improvements on the performance of this operation via
feedback of measurements of the control field at the NMR sample. Owing
to the practicality and simplicity of the protocol, we have adopted it
as a method of performing single parameter calibration of shaped
pulses. This scheme can be extended to an \emph{in situ} pulse design
protocol, in which individual parameters can be optimized iteratively
without assumptions about experimental imperfections. While it is
essential to have access to reliable single-qubit operations before
using the twirl protocol, the reliability of such operations can be
certified via randomized benchmarking, as we have demonstrated.

\appendix 

\section{Appendix A: Experimental procedure with example \label{appendixa}}

\begin{figure}
  \includegraphics[width=.5\textwidth]{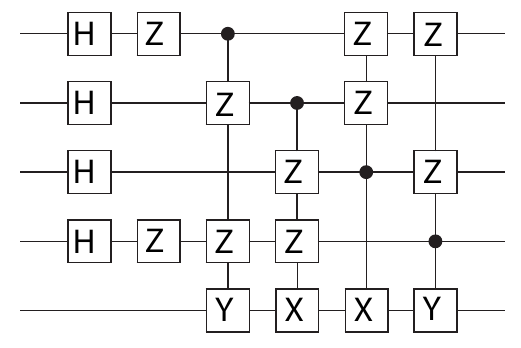}
  \caption{Encoding network for the [[5,1,3]] stabilizer code~\cite{Got97a}.\label{fig:enc5}}
\end{figure}

In this section, we describe in some detail the proposed protocol for
estimating the average fidelity of a noisy implementation $\tU$ to an
ideal $n$-qubit Clifford gate $\U$. For illustrative purposes, we use
the encoding circuit~\cite{Got97a} of the five-qubit
code~\cite{LMPZ96} (shown in Fig.~\ref{fig:enc5}) as a running
example of the unitary process to be certified. As mentioned in the
main text, this protocol can be viewed as a variation on a
local-Cliffords-and-permutations twirling scheme described in
ref.~\cite{ESM+07}, and detailed in ref.~\cite{Silva08} as the parity
monitoring protocol.

In order to estimate the average fidelity between a physical
implementation $\tU$ and the ideal encoding circuit $\U$, one can use
the following procedure.

\begin{enumerate}
\item Offline, for each $w\in{1,\cdots,n}$:

\begin{enumerate}

\item Choose an operator $\op{M}_k$ with weight $w$ from the Pauli
  group on $n$-qubits. The overall sign of the Pauli operator should
  also be chosen uniformly at random. This is a tensor product of $n$
  single-qubit Pauli operators, where $n-w$ of them are the
  identity. This step amounts to picking a random string
  of $2w+1$ bits, and choosing $w$ qubits at random on which to act
  with the non-identity Paulis.

\item For the running example, one such choice for, say, $w=3$ would
  be $\op{M}_{1} = \MI$.

\item Repeat this procedure $k_w=O(1/\epsilon^2)$ times in order
  to achieve a final accuracy of $\epsilon$~\cite{ESM+07}.

\end{enumerate}

\item In the laboratory:
\begin{enumerate}

\item For each choice of $\op{M}_k$, prepare a state $\op{\rho}_k$ such that 
\begin{equation}
r_k:=\langle{\op{M}_k}\rangle_{\op{\rho}_k} = \tr \op{\rho}_k \op{M}_k\ne 0\,.
\end{equation}
This, for example, can always be achieved by applying local Cliffords
to the state $\ket{0}^{\otimes n}$.  For $\op{M}_1=\MI$, one choice of local
operations which achieves this is $\op{C}_1 = \op{I}\otimes \op{P}\op{H}\otimes \op{I}
\otimes \op{H} \otimes \op{I} $, where
$\op{H}={1\over\sqrt{2}}\left(\begin{smallmatrix}1 & 1\\ 1 & -1\end{smallmatrix}\right)$ 
and $\op{P}=\left(\begin{smallmatrix}1 & 0\\ 0 &
    i\end{smallmatrix}\right)$, 
resulting in $r_1=1$.

\item Apply the noisy implementation $\tU$ to the prepared state
  $\op{\rho}_k$

\item Measure the expectation value
\begin{equation}
\begin{split}
  t_k:=
  &\expect{f(\op{M}_{k},C_i,\U)}{\tU(\op{\rho}_k)} \\
  = &\expect{\U(C_i\op{M}_{k}C_i^\dagger)}{\tU(\op{\rho}_k)} \\
  = &\tr\left[C_i^\dagger{\mathcal E}(C_i\ \ketbra{0}{0}^{\otimes n}\ C_i^\dagger)C_i\op{M}_k\right].
  \end{split}
\end{equation} 
If, for example, one had access only to projective measurements in the
Pauli $Z$ eigenbasis on the individual qubits, a measurement of
$f(\op{M}_{k},C_i,\U)$ can be accomplished by a basis change which is
a tensor product of single-qubit Clifford transformations. For the
running example, $f(\op{M}_{1},C_1,\U) = \MO$, so that the
transformation needed to change Pauli $Z$ measurements into this
observable would be $\op{C}_1^{'}=\op{I}\otimes \op{I}\otimes
\op{I}\otimes \op{P}\op{H}\otimes \op{H}$.

\end{enumerate}

\item The average fidelity should be estimated as follows:

\begin{enumerate}
\item For each weight $w$, $\lambda_w$ is the average of the
  ratio $t_k/r_k$ for all $\op{M}_k$ of weight $w$ ($\lambda_0$ is
  taken to be $1$).
\item $\Pr(\text{no error})$ is the inner product of $\lambda_w$ and the
  first row of $\Omega^{-1}$, which is a matrix described in
  refs.~\cite{ESM+07,Silva08}. This results in
\begin{align}
\Pr(\text{no error}) & = \sum_{w=0}^n {3^{w} {n\choose w}\over 4^n}~\lambda_w .
\end{align}

\item The average fidelity $\overline{F}$ between $\U$ and $\tU$ is
  finally given by
\begin{equation}
\overline{F} = {2^n \Pr(\text{no error}) + 1 \over 2^n + 1}
\end{equation}

\end{enumerate}
\end{enumerate}

\section{Appendix B: Software}

A simple script which automates the computation of the transformed
Pauli operators given some Clifford operation can be found at 
\url{http://github.com/marcusps/TransPauli}.

\textbf{Acknowledgements --} We thank J. Emerson, A. Blais and
J. Gambetta for comments on the manuscript.  This work was funded by
the Natural Sciences and Engineering Research Council (NSERC) of
Canada.

\bibliography{ref.bib}

\end{document}